\documentclass[a4paper]{article}

\usepackage{amsmath,amssymb,amsthm,amsfonts}
\usepackage{graphicx,epsfig,color}
\usepackage[margin=3cm]{geometry}
\usepackage{hyperref}
\numberwithin{equation}{section}

\newcommand{\vect}[1]{\boldsymbol{#1}}

\begin{document}
\title{Models of light--like charges with non--geodesic world lines}
\date{20 August 2019}
\author{
  C. G. B\"ohmer\thanks{E-mail : c.boehmer@ucl.ac.uk}\\
  Department of Mathematics, University College London,\\
  Gower Street, London WC1E 6BT, UK \\[1ex]
  and \\[1ex]
  P. A. Hogan\thanks{E-mail : peter.hogan@ucd.ie}\\
  School of Physics, University College Dublin, \\
  Belfield, Dublin 4, Ireland
}

\maketitle

\begin{abstract}
Massless particles in General Relativity move with the speed of light, their trajectories in spacetime are described by null geodesics. This is independent of the electrical charge of the particle being considered, however, the charged light--like case is less well understood. Starting with the Maxwell field of a charged particle having a light--like geodesic world line in Minkowskian space--time we construct the Maxwell field of such a particle having a non--geodesic, light--like world line. The necessary geometry in the neighbourhood of an arbitrary null world line in Minkowskian space--time is described and properties of the resulting electromagnetic field are discussed. The electromagnetic field obtained represents a light--like analogue of the Li\'enard--Wiechert field, which generalises the Coulomb field of a charge having a time--like geodesic world line to the field of a charge having an accelerated world line.
\end{abstract}

\section{Introduction}

It is an interesting and noteworthy fact that a charged particle travelling with the speed of light has yet to be observed in nature. There are no field theoretical considerations which would in principle contradict the existence of such a particle. Nevertheless, in exploring the limits of classical electrodynamics it is intriguing to seek models of such particles. This paper demonstrates explicitly that more than one model exists and it will require further knowledge of the properties of such particles, if and when they are observed, to distinguish between them.

A particle with electrical charge $e$ having a time--like geodesic world line in Minkowskian space--time has a Maxwell field described by the Coulomb solution of the vacuum Maxwell field equations. If the world line is not a geodesic, i.e.~if the particle has non--vanishing 4--acceleration, then its Maxwell field is described by the Li\'enard--Wiechert solution of the vacuum Maxwell field equations. The Li\'enard--Wiechert electromagnetic field has the property that near the world line of the charged particle it resembles the Coulomb field and far from the world line it describes the electromagnetic radiation produced by the acceleration of the charge.

The Li\'enard--Wiechert 4--potential, when written in rectangular Cartesian coordinates and time, has the property of being proportional to the 4--velocity of the particle, modulo a gauge transformation. Many years ago Synge~\cite{S1} looked for a Maxwell field of an accelerated light--like charge by choosing a 4--potential proportional to the null tangent to the world line of the charge. This resulted in an electromagnetic field which did not contain an analogue of the Coulomb part of the Li\'enard--Wiechert field but described electromagnetic radiation produced by the accelerated charge. If the charge has a null geodesic world line then the electromagnetic field vanishes. Charged gyratons which are models of massless charged particles with spin have been studied in~\cite{FRO}. This result suggests that it might also be possible to study charged light--like particles using a mainly geometrical approach. This is the aim of the present paper. We seek to construct a model of an accelerated light--like charge which incorporates an analogue of the Coulomb part of the Li\'enard--Wiechert field and an analogue of the radiation part of the Li\'enard--Wiechert field. Properties of hypothetical charged particles moving with the speed the of light were already studied as early as the 1940s and were based on an entirely classical treatment, see in particular~\cite{WR1,WR2}.

We begin in Section~\ref{sec:cou} by describing the electromagnetic field of a charged particle having a null geodesic world line. This is a light--like analogue of the Coulomb field. The result is a spin--off from the Robinson--Trautman~\cite{RT1,RT2} solutions of the vacuum Einstein--Maxwell field equations. It exploits the idea of using null geodesics to set up a coordinate system which is ideally suited to the description of light--like particle. In Section~\ref{sec:geo} we develop the geometry associated with a non--geodesic, light--like world line in Minkowskian space--time. A byproduct of this study is to establish the existence of a parameter along the world line which is unique up to a linear transformation and which specialises to an affine parameter if the world line is a null geodesic. This is important because, in contradistinction to the time--like case, we do not have the arc length available to us as a parameter along the world line in the light--like case. The existence of such a parameter is one of the key ingredients of the final construction. Consequently, a model of the electromagnetic field of a light--like charge, in the form of a solution of Maxwell's vacuum field equations on Minkowskian space--time is derived in Section~\ref{sec:acccharge} and some properties of the model are discussed in Section~\ref{sec:prop}. We conclude our work with discussions in the final section.

\section{Light--like analogue of the Coulomb field}
\label{sec:cou}

All topics under consideration in this paper are in the context of Minkowskian space--time. The Minkowskian line element in rectangular Cartesian coordinates $X^i=(T,X, Y, Z)$ reads
\begin{align}
  \label{1}
  ds^2 = \eta_{ij}\,dX^i\,dX^j =
  (dT)^2-(dX)^2-(dY)^2-(dZ)^2\ .
\end{align}
We are working with signature $(+,-,-,-)$. The world line in Minkowski space of a point charge giving rise to the Coulomb field is a time--like geodesic. For the light--like analogue of the Coulomb field the world line of the charge will be a null geodesic. We take this null geodesic to have parametric equations
\begin{align}
  \label{2}
  X^i(u) = u\,v^i \quad {\rm with}\quad v^i=(1, 0, 0, 1)\ ,
\end{align}
where $v^i$ is a null vector because of $\eta_{ij}\,v^i\,v^j=v_i\,v^i=0$. The quantity $u$ is an affine parameter along the null geodesic with tangent $v^i$ and we can take $u \in \mathbb{R}$.

We will now introduce a new set of coordinates of the position 4--vector of a point in Minkowski space relative to this null geodesic as follows
\begin{align}
  \label{3}
  X^i = u\,v^i + r\,k^i\ ,
\end{align}
and we choose $k^i$ to satisfy
\begin{align}
  \label{4}
  k^i\,k_i=0 \quad {\rm and}\quad k^i\,v_i=+1\ .
\end{align}
Hence, the world line (\ref{2}) corresponds to $r=0$ and we shall take $0\leq r<+\infty$. This particular construction will prove very useful in the following as it is intimately tied to the geometry of a particle moving at the speed of light. 

The vector $k^i$ is null and normalised relative to $v^i$ which means it can be parametrised by two real parameters $\xi \in \mathbb{R}$ and $\eta \in \mathbb{R}$. They determine the \emph{direction} of $k^i$ in space--time, we can write this vector as
\begin{align}
  \label{5}
  k^i=\left (\frac{1}{2}(\xi^2+\eta^2+1),\xi, \eta, \frac{1}{2}(\xi^2+\eta^2-1) \right)\ .
\end{align}
For sufficiently large values of $\xi$ and $\eta$ we write $\zeta = \sqrt{\xi^2 + \eta^2}$ and only keep terms in the highest power in $\zeta$. This gives
\begin{align}
  \label{5.1}
  k^i \rightarrow \frac{\zeta^2}{2} (1,0,0,1) = \frac{\zeta^2}{2} v^i\ ,
\end{align}
for large $\zeta$. This means that $k^i$ points in the direction of $v^i$ in (\ref{2}) for large $\xi$ and $\eta$.

Let us now consider (\ref{3}) as a coordinate transformation between the original Cartesian coordinates and the new coordinates $x^i=(u, \xi, \eta, r)$. Writing this transformation out explicitly gives
\begin{align}
  \label{6}
  T &= u + \frac{r}{2}(\xi^2+\eta^2+1)\ , \\
  Z &= u + \frac{r}{2}(\xi^2+\eta^2-1)\ , \\
  X &= r\,\xi \ , \quad Y=r\,\eta\ .
  \label{6.1}
\end{align}

Substituting (\ref{6})--(\ref{6.1}) into the Minkowski line element (\ref{1}) results in
\begin{align}
  \label{7}
  ds^2 = 2\,du\,dr - r^2(d\xi^2+d\eta^2) =
  2\,\vartheta^0\,\vartheta^3 -(\vartheta^1)^2-(\vartheta^2)^2 \ ,
\end{align}
with the basis 1--forms $\vartheta^0, \vartheta^1, \vartheta^2, \vartheta^3$ given by
\begin{align}
  \label{8}
  \vartheta^0=du\ ,\quad
  \vartheta^1=r\,d\xi\ ,\quad
  \vartheta^2=r\,d\eta\ , \quad
  \vartheta^3=dr\ .
\end{align}

As the potential 1--form due to a particle of charge $e$ (which we assume to be constant) with world line $r=0$, the light--like analogue of the Coulomb potential, we take
\begin{align}
  \label{9}
  A = \frac{e}{r}\,\vartheta^0 = \frac{e}{r}\,du\ ,
\end{align}
where we emphasise that $u$ was the affine parameter along the null geodesic. The corresponding candidate for the Maxwell field due to this charged particle is the exterior derivative of $A$ resulting in the 2--form
\begin{align}
  \label{10}
  F = dA = \frac{e}{r^2}\,du\wedge dr =
  \frac{e}{r^2}\,\vartheta^0\wedge\vartheta^3\ .
\end{align}
The Hodge dual of $F$ is the 2--form ${}^*F$ given by
\begin{align}
  \label{11}
        {}^*F=\frac{e}{r^2}\,\vartheta^1\wedge\vartheta^2=e\,d\xi\wedge d\eta\ ,
\end{align}
from which it immediately follows that Maxwell's vacuum field equations
\begin{align}
  \label{12}
  d{}^*F=0\ ,
\end{align}
are satisfied. Therefore, the potential 1--form (\ref{9}) describes the Maxwell field of a light--like particle and can be seen as the analogue of the Coulomb field of a time--like particle. The Maxwell field of such a light--like particle is a spin--off of the charged Robinson--Trautman fields~\cite{RT1,RT2} which are solutions of the vacuum Einstein--Maxwell field equations. It is worth pointing out that the entire construction of this solution was based on exploiting the inherent geometry of Minkowski space in the presence of particles moving at the speed of light.

\section{Geometry associated with an accelerated light--like world line}
\label{sec:geo}

\begin{figure}[!htb]
  \centering
  \includegraphics[trim={0 2.5cm 0 0.5cm},clip,width=0.8\textwidth]{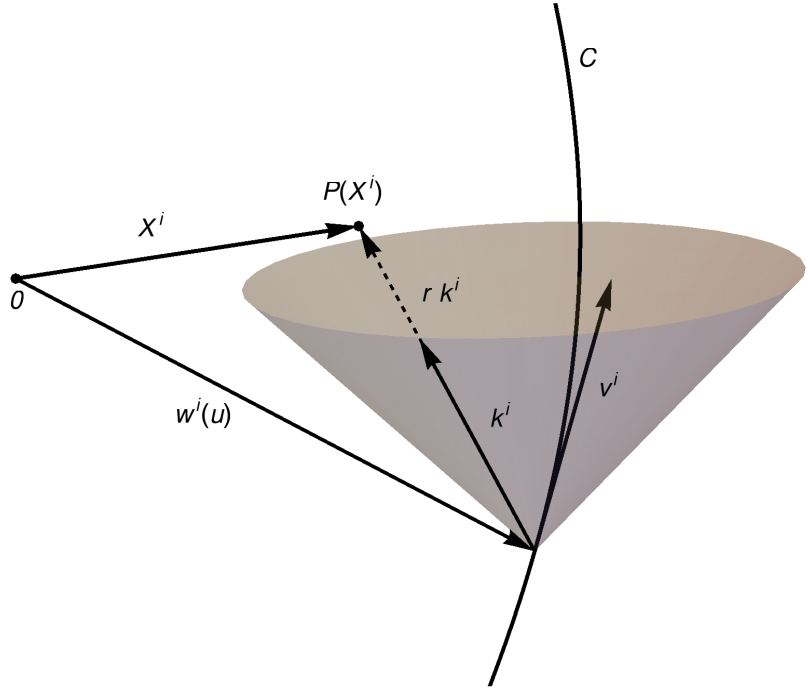}
    \caption{The light--like world line (curve) is denoted by $C$, its tangent vector is $v^i$. $k^i$ denotes the null vector and $r$ is the `distance' to the worldline. $X^i$ is the position vector of a point $P$ relative to the origin $0$.}
\label{fig_coord_nobar}
\end{figure}

We generalise the choice of coordinates (\ref{3}) to a position 4--vector in Minkowskian space--time relative to an arbitrary light--like world line with parametrisation $X^i=w^i(u)$. The functions $w^i(u)$ were introduced to make the notation clearer and to distinguish with the previous case. The tangent vector to this curve is given by $v^i(u)=dw^i/du$ and satisfies $v_i\,v^i=0$, as before. Moreover, we introduce the acceleration $a^i(u)=dv^i/du$ which satisfies $a_i\,v^i=0$. This follows from differentiating $v_i\,v^i=0$ with respect to the parameter $u$. In general $a^i\neq 0$ and so the light--like world line is not necessarily a geodesic and the particle having this world line as its history will be said to be accelerated. The parameter $u$ along the world line, for which $u \in \mathbb{R}$, is unspecified and we will exploit this fact later. We replace (\ref{3}) by the more general
\begin{align}
  \label{13}
  X^i=w^i(u)+r\,k^i\ ,
\end{align}
with $k_i\,k^i=0$ and $k_i\,v^i=+1$. Hence the light--like world line corresponds to $r=0$. This setup is visualised in Fig.~\ref{fig_coord_nobar}.

As in the previous discussion, we parametrise the direction of $k^i$ with the two real parameters $x \in \mathbb{R}$ and $y \in \mathbb{R}$ and write
\begin{align}
  \label{14}
  -P_0\,k^i=\left(-1-\frac{1}{4}(x^2+y^2),x, y, 1-\frac{1}{4}(x^2+y^2)\right)\ ,
\end{align}
for some function $P_0(x, y, u)$. This function $P_0$ is determined by the normalisation of the vector $k^i$ to be $v_i\,k^i=+1$ which means we have
\begin{align}
  \label{15}
  P_0 = \{1+\frac{1}{4}(x^2+y^2)\}\,v^0(u) + x\,v^1(u)+y\,v^2(u)+\{1-\frac{1}{4}(x^2+y^2)\}\,v^3(u) \ .
\end{align}
One should note that the components of the velocity are functions of the parameter $u$ only, the spatial dependence of $P_0$ enters entirely through the components of the vector $k^i$. A direct calculation shows that $P_0$ satisfies
\begin{align}
  \label{16}
  P_0^2\left (\frac{\partial^2}{\partial x^2}+\frac{\partial^2}{\partial y^2}\right )\log P_0 =
  \Delta \log P_0 = v^i\,v_i=0\ ,
\end{align}
where $\Delta$ stands for the covariant Laplacian on the 2-surface with line element $P_0^{-2}(dx^2 + dy^2)$.

Contracting  (\ref{14}) with the acceleration vector yields the relation
\begin{align}
  \label{17}
  P_0\,a_i\,k^i = \{1+\frac{1}{4}(x^2+y^2)\}\,a^0(u) + x\,a^1(u) + y\,a^2(u) +
  \{1-\frac{1}{4}(x^2+y^2)\}\,a^3(u)\ ,
\end{align}
and thus one can deduce that $P_0$ satisfies
\begin{align}
  \label{18}
  \frac{\partial}{\partial u}\log P_0 = a_i\,k^i =: h_0\ ,
\end{align}
which defines the new function $h_0$.

At this point we shall assume that $v^0-v^3\neq 0$. If $v^0=v^3$ then, since $v^i$ is a null vector, we have $v^1=0=v^2$ and also $a^i$ must be in the same direction (the $(T,Z)$--plane) as $v^i$ and so $r=0$ is a null geodesic and we are led back to the situation discussed in Section~\ref{sec:cou}. So assuming from now on that $v^0-v^3 \neq 0$ we can rewrite (\ref{15}) in the useful form
\begin{align}
  \label{19}
  P_0=\frac{(v^0-v^3)}{4}
  \left\{
  \left (x+\frac{2\,v^1}{v^0-v^3}\right )^2+\left (y+\frac{2\,v^2}{v^0-v^3}\right )^2
  \right\}\ .
\end{align}
Note that this form of $P_0$ again shows directly that $\Delta \log P_0 = 0$. Following on from the previous construction, we introduce $(u, x, y, r)$ as coordinates, instead of $X^i$ which are related to the former by (\ref{13}). The Minkowski line element (\ref{1}) now takes the form
\begin{align}
  \label{20}
  ds^2 = 2\,du\,dr - 2\,h_0\,r\,du^2 - P_0^{-2}(dx^2+dy^2) \ .
\end{align}
The form of $P_0$ given in (\ref{19}) suggests a coordinate transformation from $x, y$ to $\xi, \eta$ given by
\begin{align}
  \xi &= \frac{1}{P_0} \left(x+\frac{2\,v^1}{v^0-v^3}\right),
  \label{21}\\
  \eta &= \frac{1}{P_0} \left(y+\frac{2\,v^2}{v^0-v^3}\right).
  \label{22}
\end{align}
When these new coordinates are substituted into (\ref{20}) the result is the line element
\begin{align}
  \label{23}
  ds^2 = 2\,du\,dr-2\,h_0\,r\,du^2 -
  r^2\left\{
  \left (d\xi+\frac{\partial q}{\partial\eta}\,du\right )^2+
  \left(d\eta+\frac{\partial q}{\partial\xi}\,du\right )^2
  \right\}\ ,
\end{align}
with the function $q(\xi, \eta, u)$ given by
\begin{align}
  \label{24}
  q(\xi, \eta, u) = -\frac{1}{6}(\eta^3 A^1 + \xi^3 A^2) +
  \xi \eta \left(\frac{1}{2}(\xi A^1 + \eta A^2) + \frac{a^0-a^3}{v^0-v^3}\right)
\end{align}
and
\begin{align}
  A^1 = a^1-\left (\frac{a^0-a^3}{v^0-v^3}\right )v^1\ , \qquad
  A^2 = a^2-\left (\frac{a^0-a^3}{v^0-v^3}\right )v^2\ .
  \label{25}
\end{align}
The null vector field $k^i$ in (\ref{14}) written in terms of the parameters $\xi, \eta$ instead of $x, y$ can now be written in the form
\begin{align}
  \label{27}
  k^i= \zeta^i - \frac{1}{2} (\zeta_k\,\zeta^k)\,v^i\quad {\rm with}\quad
  \zeta^i=\left (\frac{1-\xi\,v^1-\eta\,v^2}{v^0-v^3},-\xi, -\eta, \frac{1-\xi\,v^1-\eta\,v^2}{v^0-v^3}\right )\ .
\end{align}
Since $\zeta_i\,\zeta^i=-(\xi^2+\eta^2) = -\zeta^2$ we see that for large values of $\xi$ and $\eta$ the null vector $k^i$ points in the direction of the tangent $v^i$ to the world line $r=0$, similar to the previous result~(\ref{5.1}). We also note that $q$ is a harmonic function and thus
\begin{align}
  \label{28}
  \Delta q=\frac{\partial^2q}{\partial\xi^2}+\frac{\partial^2q}{\partial\eta^2}=0\ ,
\end{align}
and from (\ref{27}) we have
\begin{align}
  \label{29}
  h_0=a_i\,k^i=\xi\,A^1+\eta\,A^2+\left (\frac{a^0-a^3}{v^0-v^3}\right ) =
  \frac{\partial^2q}{\partial\xi\partial\eta}\ .
\end{align}

At this point the parameter $u$ along the light--like world line $r=0$ is unspecified. It is useful to specify it up to a linear transformation as follows: Start with the coordinate transformation
\begin{align}
  \label{30}
  \bar\xi=\mu\,\xi, \quad
  \bar\eta=\mu\,\eta, \quad
  \bar r=\mu^{-1}r, \quad
  \bar u=\bar u(u)\ ,
\end{align}
with
\begin{align}
  \label{31}
  \mu^{-1}\frac{d\mu}{du}=\frac{a^0-a^3}{v^0-v^3}
  \quad {\rm and}\quad
  \frac{d\bar u}{du}=\mu\ .
\end{align}
We note from (\ref{31}) that if we take 
\begin{align}
  \label{32}
  \bar u=\int(v^0-v^3)\,du\ ,
\end{align}
then $\bar u$ is unique up to a linear transformation $\bar u\rightarrow c_1\,\bar u+c_2$ where $c_1, c_2$ are two real constants. If we let
\begin{align}
  \label{33}
  A^i=a^i-\left (\frac{a^0-a^3}{v^0-v^3}\right )\,v^i\ ,
\end{align}
then the cases $i=1$ and $i=2$ are given in (\ref{25}). If the world line $r=0$ is a null geodesic then, in general, $a^i=\lambda(u)\,v^i$ for some function $\lambda(u)$ and $A^i=0$. The change of parameter $u$ along the world line $r=0$ described by (\ref{31}) results in 
\begin{align}
  \label{34}
  v^i=\bar v^i\,\mu\quad {\rm and}\quad
  a^i=\bar v^i\,\mu\,\left (\frac{a^0-a^3}{v^0-v^3}\right )+\bar a^i\,\mu^2\ ,
\end{align}
where $\bar v^i=dw^i/d\bar u$ and $\bar a^i=d\bar v^i/d\bar u$. When this is substituted into (\ref{33}) we find that
\begin{align}
  \label{35}
  A^i=\mu^2\bar a^i\ ,\end{align}
and so we have the result that
\begin{align}
  \label{36}
  a^i=\lambda(u)\,v^i\quad \Rightarrow\quad \bar a^i=0\ .
\end{align}
Hence we see an important property of the parameter $\bar u$, namely, if $r=0$ is a geodesic then $\bar u$ is an affine parameter along it, see~\cite{Kaz} for a similar discussion.

Now when the coordinate transformation (\ref{30}) with (\ref{31}) is applied to the line element (\ref{23}) the result is 
\begin{align}
  \label{37}
  ds^2=2\,d\bar u\,d\bar r-2\,\bar h_0\,\bar r\,d\bar u^2 -
  \bar r^2\left\{
  \left (d\bar\xi+\frac{\partial \bar q}{\partial\bar \eta}\,d\bar u\right )^2+
  \left (d\bar\eta+\frac{\partial \bar q}{\partial\bar\xi}\,d\bar u\right )^2
  \right\}\ ,
\end{align}
with
\begin{align}
  \label{38}
  \bar q(\bar\xi, \bar\eta, \bar u)=
  -\frac{1}{6}(\bar{\eta}^3 \bar{a}^1 + \bar{\xi}^3 \bar{a}^2) +
  \frac{1}{2}\xi \eta (\bar{\xi} \bar{a}^1 + \bar{\eta} \bar{a}^2)
\end{align}
and
\begin{align}
  \label{39}
  \bar h_0=\frac{\partial^2\bar q}{\partial\bar\xi\partial\bar\eta}\ .
\end{align}
Finally we note that since (\ref{33}) implies that $A^0=A^3$ we have from (\ref{35}) that $\bar a^0=\bar a^3$ because $\mu\neq 0$. These statements rely on our assumptions that $v^0-v^3\neq 0$. This, together with the orthogonality of $\bar v^i$ and $\bar a^i$, allows us to conlcude that
\begin{align}
  \bar a^0 = \bar a^3 =
  \frac{\bar v^1}{\bar v^0-\bar v^3}\bar a^1 + \frac{\bar v^2}{\bar v^0-\bar v^3}\bar a^2 \ .
\end{align}
We are now able to apply these result to the study of an accelerated light--like charge by following the exact same geometrical setup.

\section{Accelerated Light--Like Charge}
\label{sec:acccharge}

Guided by the analogue of the Coulomb solution of Maxwell's equations described in Section~\ref{sec:cou}, the work of Robinson and Trautman~\cite{RT1,RT2} on solutions of the vacuum Einstein--Maxwell field equations, and requiring the solution of Maxwell's equations for the electromagnetic field of an accelerating light--like charge to specialise to the case of an unaccelerated light--like charge (\ref{9}), we look for a potential 1--form to describe the Maxwell field of an accelerated light--like charge of the form
\begin{align}
  \label{40}
  A = e\left (\frac{1}{\bar r}+G(\bar\xi, \bar\eta, \bar u)\right )d\bar u\ .
\end{align}
The aim is to specify the function $G(\bar\xi, \bar\eta, \bar u)$ so that the Einstein--Maxwell equations are satisfied. Following on from the previously introduced basis 1--forms of the line element (\ref{37}), we set
\begin{align}
  \label{41}
  \bar\vartheta^0 &= d\bar u\ , \quad \bar\vartheta^3=d\bar r-\bar h_0\,\bar r\,d\bar u\ ,\\
  \bar\vartheta^1 & =\bar r\left (d\bar\xi+\frac{\partial\bar q}{\partial\bar\eta}d\bar u\right ),\quad
  \bar\vartheta^2 = \bar r\left (d\bar\eta+\frac{\partial\bar q}{\partial\bar\xi}d\bar u\right ). 
\end{align}
The candidate for Maxwell 2--form is the exterior derivative of (\ref{40}) which reads
\begin{align}
  F &= dA=
  -\frac{e}{\bar r^2}\,d\bar r\wedge d\bar u
  +e\,\frac{\partial G}{\partial\bar\xi}\,d\bar\xi\wedge d\bar u
  +e\,\frac{\partial G}{\partial\bar\eta}\,d\bar\eta\wedge d\bar u\ ,
  \nonumber\\
  &= -\frac{e}{\bar r}\left(
  \frac{1}{\bar r}\,\bar\vartheta^3
  -\frac{\partial G}{\partial\bar\xi}\,\bar\vartheta^1
  -\frac{\partial G}{\partial\bar\eta}\,\bar\vartheta^2
  \right)\wedge\bar\vartheta^0\ .
\label{42}
\end{align}
The Hodge dual of this 2--form becomes
\begin{align}
  {}^*F & = \frac{e}{\bar r}\left(
  \frac{1}{\bar r}\,\bar\vartheta^1\wedge\bar\vartheta^2
  +\frac{\partial G}{\partial\bar\xi}\,\bar\vartheta^2\wedge\bar\vartheta^0
  -\frac{\partial G}{\partial\bar\eta}\,\bar\vartheta^1\wedge\bar\vartheta^0
  \right)
  \nonumber\\
  &=e\,d\bar\xi\wedge d\bar\eta+
  e\,\left (\frac{\partial\bar q}{\partial\bar\xi}-\frac{\partial G}{\partial\bar\eta}\right )\,d\bar\xi\wedge d\bar u-e\,\left (\frac{\partial\bar q}{\partial\bar\eta}-\frac{\partial G}{\partial\bar\xi}\right )\,d\bar\eta\wedge d\bar u\ ,
  \label{43}
\end{align}
from which one immediately arrives at
\begin{align}
  \label{44}
  d{}^*F=e\,\left(
  -2\frac{\partial^2\bar q}{\partial\bar\xi\partial\bar\eta}
  +\frac{\partial^2G}{\partial\bar\xi^2}
  +\frac{\partial^2G}{\partial\bar\eta^2}
  \right )\,
  d\bar\xi\wedge d\bar\eta\wedge d\bar u\ .
\end{align}
Therefore Maxwell's vacuum field equations $d{}^*F=0$ imply that $G$ must satisfy
\begin{align}
  \label{45}
  \Delta G = \frac{\partial^2G}{\partial\bar\xi^2}+\frac{\partial^2G}{\partial\bar\eta^2} =
  2\frac{\partial^2\bar q}{\partial\bar\xi\partial\bar\eta}=2\,\bar h_0\ .
\end{align}
Since $\bar q$, given by (\ref{38}), is also a harmonic function satisfying $\Delta\bar h_0=0$, it follows that $G$ satisfies the biharmonic equation
\begin{align}
  \label{46}
  \Delta\Delta G=0\ .
\end{align}

It is well known that the general solution of this equation is (a proof due to A. Schild is given in~\cite{SHC})
\begin{align}
  \label{47}
  G(\bar\xi, \bar\eta, \bar u)=
  {\rm Re}\{f(\bar\xi+i\bar\eta, \bar u)+(\bar\xi-i\bar\eta)\,F(\bar\xi+i\bar\eta, \bar u)\}\ ,
\end{align}
where $f, F$ are arbitrary analytic functions of $\bar\xi+i\bar\eta$. This means
\begin{align}
  \label{48}
  f(\bar\xi+i\bar\eta, \bar u)=U(\bar\xi, \bar\eta, \bar u)+iV(\bar\xi, \bar\eta, \bar u)\quad
  {\rm with}\quad
  \frac{\partial U}{\partial\bar\xi}=\frac{\partial V}{\partial\bar\eta}\ ,\ 
  \frac{\partial U}{\partial\bar\eta}=-\frac{\partial V}{\partial\bar\xi}\ ,
\end{align}
and
\begin{align}
  \label{49}
  F(\bar\xi+i\bar\eta, \bar u)=W(\bar\xi, \bar\eta, \bar u)+iS(\bar\xi, \bar\eta, \bar u) \quad
  {\rm with}\quad
  \frac{\partial W}{\partial\bar\xi}=\frac{\partial S}{\partial\bar\eta}\ ,\ 
  \frac{\partial W}{\partial\bar\eta}=-\frac{\partial S}{\partial\bar\xi}\ .
\end{align}
Hence (\ref{47}) reads
\begin{align}
  \label{50}
  G(\bar\xi, \bar\eta, \bar u)=U+\bar\xi\,W+\bar\eta\,S\ .
\end{align}
We note in passing that we could equally well have used the imaginary part of $f+(\bar\xi+i\bar\eta)\,F$ for $G$ in (\ref{47}). Clearly not all solutions of (\ref{46}) are solutions of (\ref{45}) and so substituting (\ref{50}) into (\ref{45}) yields
\begin{align}
  \label{51}
  \Delta G = 2\,\left (\frac{\partial W}{\partial\bar\xi}+\frac{\partial S}{\partial\bar\eta}\right )=
  4\,\frac{\partial W}{\partial\bar\xi}=4\,\frac{\partial S}{\partial\bar\eta}=
  2\frac{\partial^2\bar q}{\partial\bar\xi\partial\bar\eta}\ .
\end{align}
From this we have
\begin{align}
  \label{52}
  W=\frac{1}{2}\,\frac{\partial\bar q}{\partial\bar\eta}+\alpha(\bar\eta)\quad
  {\rm and}\quad
  S=\frac{1}{2}\,\frac{\partial\bar q}{\partial\bar\xi}+\beta(\bar\xi)\ ,
\end{align}
where $\alpha, \beta$ are functions of integration. But
\begin{align}
  \label{53}
  0=\frac{\partial W}{\partial\bar\eta}+\frac{\partial S}{\partial\bar\xi}=\frac{1}{2}\,\Delta\bar q+\frac{d\alpha}{d\bar\eta}+\frac{d\beta}{d\bar\xi}=\frac{d\alpha}{d\bar\eta}+\frac{d\beta}{d\bar\xi}\ ,
\end{align}
since $\bar q$ is a harmonic function, and hence we must have
\begin{align}
  \label{54}
  \frac{d\alpha}{d\bar\eta}=C_1=-\frac{d\beta}{d\bar\xi}
  \quad \Rightarrow\quad
  \alpha(\bar\eta)=C_1\,\bar\eta+C_2\ ,\ \beta(\bar\xi)=-C_1\,\bar\xi+C_3\ ,
\end{align}
where $C_1$ is a separation constant and $C_2, C_3$ are constants of integration. Substituting (\ref{52}) with (\ref{54}) into (\ref{50}) gives
\begin{align}
  \label{55}
  G =\underbrace{\frac{1}{2}\,\left (\bar\xi\,\frac{\partial\bar q}{\partial\bar\eta}+\bar\eta\,\frac{\partial\bar q}{\partial\bar\xi}\right )}_{\text{field of particle}} +
  \underbrace{U+C_2\,\bar\xi+C_3\,\bar\eta}_{\text{spherical EM waves}} =
  G_{\text{particle}} + G_{\text{spherical}}\ .
\end{align} 
The last three terms $G_{\text{spherical}}$ here constitute an arbitrary harmonic function, clearly $\Delta G_{\text{spherical}} = 0$. When substituted into the Maxwell field (\ref{42}) this harmonic function describes spherical electromagnetic waves which are independent of the accelerated light--like particle and will therefore be excluded henceforth. The electromagnetic field of the light--like particle is described simply by (\ref{42}) with
\begin{align}
  \label{56}
  G_{\text{particle}}=\frac{1}{2}\,\left (\bar\xi\,\frac{\partial\bar q}{\partial\bar\eta}+
  \bar\eta\,\frac{\partial\bar q}{\partial\bar\xi}\right ) =
  \frac{1}{4}\,(\bar\xi^2+\bar\eta^2)\{\bar a^1(u)\,\bar\xi+\bar a^2(u)\,\bar\eta\}\ .
\end{align}
If the world line of the particle is a null geodesic then $\bar a^i(\bar u)=0$ and the Maxwell field 
of the accelerated light--like particle specialises to the case described in Section~\ref{sec:cou}.

\section{Properties of the solution}
\label{sec:prop}

To obtain a useful comparison between the Maxwell field constructed from (\ref{42}) and (\ref{56}) and the Maxwell field of a point charge with a time--like world line we proceed as follows: Start by substituting the transformations (\ref{30}) into the position vector (\ref{13}) which becomes
\begin{align}
  \label{57}
  X^i=w^i(\bar u)+\bar r\,\bar k^i\ ,
\end{align}
where $\bar k^i$ is given by 
\begin{align}
  \label{58}
  \bar k^i=\bar q^i-\frac{1}{2}(\bar q_k\,\bar q^k)\,\bar v^i
  \quad {\rm with}\quad
  \bar q^i=\left(
  \frac{1-\bar\xi\,\bar v^1-\bar\eta\,\bar v^2}{\bar v^0-\bar v^3},
  -\bar\xi,
  -\bar\eta,
  \frac{1-\bar\xi\,\bar v^1-\bar\eta\,\bar v^2}{\bar v^0-\bar v^3}
  \right )\ .
\end{align}
Geometrically, this is the same setup as shown in Fig.~\ref{fig_coord_nobar} with all quantities replaced by their barred counterparts. From this $\bar k^i$ we calculate the useful formulae
\begin{align}
  \label{59}
  \frac{\partial\bar k^i}{\partial\bar\xi}\,\frac{\partial\bar k_i}{\partial\bar\xi}=
  \frac{\partial\bar k^i}{\partial\bar\eta}\,\frac{\partial\bar k_i}{\partial\bar\eta}=-1\ ,\quad
  \frac{\partial\bar k^i}{\partial\bar\xi}\,\frac{\partial\bar k_i}{\partial\bar\eta}=0\ ,
\end{align}
and
\begin{align}
  \label{60}
  \frac{\partial\bar k^i}{\partial\bar\xi}\,\frac{\partial\bar k_i}{\partial\bar u}=
  -\frac{\partial\bar q}{\partial\bar\eta}\ ,\quad
  \frac{\partial\bar k^i}{\partial\bar\eta}\,\frac{\partial\bar k_i}{\partial\bar u}=
  -\frac{\partial\bar q}{\partial\bar\xi}\ .
\end{align}
Consequently we can reaarange to find
\begin{align}
  \label{61}
  \frac{\partial\bar k^i}{\partial\bar u}=
  -\bar h_0\,\bar k^i
  + \frac{\partial\bar q}{\partial\bar\eta}\,\frac{\partial\bar k^i}{\partial\bar\xi}
  + \frac{\partial\bar q}{\partial\bar\xi}\,\frac{\partial\bar k^i}{\partial\bar\eta}\ .
\end{align}
Equation (\ref{57}) implicitly defines $\bar\xi, \bar\eta, \bar r, \bar u$ as functions of $X^i$. The gradients of these functions are obtained by first differentiating (\ref{57}) with respect to $X^j$ to find that
\begin{align}
  \label{62}
  \delta^i_j = \left (\bar v^i+\bar r\,\frac{\partial\bar k^i}{\partial\bar u}\right )\bar u_{,j} +
  \bar k^i\,\bar r_{,j}+
  \bar r\,\frac{\partial\bar k^i}{\partial\bar\xi}\,\bar\xi_{,j}+
  \bar r\,\frac{\partial\bar k^i}{\partial\bar\eta}\,\bar\eta_{,j}\ .
\end{align}
Multiplying this equation successively by $\bar k_i, \bar v_i, \partial\bar k_i/\partial\bar\xi, \partial\bar k_i/\partial\bar\eta$ and using (\ref{59}) and (\ref{60}) results in
\begin{align}
  \label{63}
  \bar u_{,j}=\bar k_j\ ,\quad \bar r_{,j}=\bar v_j+\bar r\,\bar h_0\,\bar k_j\ ,\quad
  \bar\xi_{,j}=-\frac{1}{\bar r}\frac{\partial\bar k_j}{\partial\bar\xi}-\frac{\partial\bar q}{\partial\bar\eta}\,\bar k_j\ ,\quad 
  \bar\eta_{,j}=-\frac{1}{\bar r}\frac{\partial\bar k_j}{\partial\bar\eta}-\frac{\partial\bar q}{\partial\bar\xi}\,\bar k_j\ .
\end{align}
Now the potential 1--form (\ref{40}) with (\ref{56}) is equivalent to the 4--potential in coordinates $X^i$:
\begin{align}
  \label{64}
  A_i=e\left (\frac{1}{\bar r}+G(\bar\xi, \bar\eta, \bar u)\right )\bar k_i\ .
\end{align}
The Maxwell field (\ref{42}) has components $F_{ij}=-F_{ji}$ in coordinates $X^i$ given by
\begin{align}
  F_{ij} &= A_{j,i}-A_{i,j}
  \nonumber\\
  &= \frac{e}{\bar r^2}(\bar k_i\,\bar v_j-\bar k_j\,\bar v_i) +
  \frac{e}{\bar r}\Biggl\{\frac{\partial G}{\partial\bar\xi}\left (\bar k_i\,\frac{\partial\bar k_j}{\partial\bar\xi}-\bar k_j\,\frac{\partial\bar k_i}{\partial\bar\xi}\right )
  +\frac{\partial G}{\partial\bar\eta}\left (\bar k_i\,\frac{\partial\bar k_j}{\partial\bar\eta}-\bar k_j\,\frac{\partial\bar k_i}{\partial\bar\eta}\right )\Biggr\}\ .
  \label{65}
\end{align}
The dual of this Maxwell field has components ${}^*F_{ij}=\frac{1}{2}\epsilon_{ijkl}\,F^{kl}$, with $\epsilon_{ijkl}$ the Levi--Civit\`a permutation symbol in four dimensions, and 
these components are given by
\begin{align}
  {}^*F_{ij}  = &\frac{e}{\bar r^2}\left(\frac{\partial\bar k_i}{\partial\bar\xi}\frac{\partial\bar k_j}{\partial\bar\eta}-\frac{\partial\bar k_i}{\partial\bar\eta}\frac{\partial\bar k_j}{\partial\bar\xi}\right )
  \nonumber \\ &{} +
  \frac{e}{\bar r}\Biggl\{\frac{\partial G}{\partial\bar\xi}\left (\bar k_i\,\frac{\partial\bar k_j}{\partial\bar\eta}-\bar k_j\,\frac{\partial\bar k_i}{\partial\bar\eta}\right ) - \frac{\partial G}{\partial\bar\eta}\left (\bar k_i\,\frac{\partial\bar k_j}{\partial\bar\xi}-\bar k_j\,\frac{\partial\bar k_i}{\partial\bar\xi}\right )\Biggr\}\ ,
\label{66}
\end{align}
which is equivalent to (\ref{43}). We note that
\begin{align}
  \label{67}
  F_{ij}\,\bar k^j=\frac{e}{\bar r^2}\,\bar k_i\ \ {\rm and}\ \ {}^*F_{ij}\,\bar k^j=0\ .
\end{align}

Hence the field (\ref{65}) is qualitatively similar, from an algebraic point of view, to the electromagnetic field of an accelerated point charge travelling with less than the speed of light. The $\bar r^{-2}$--part of the field is algebraically general and corresponds to the Coulomb part of the field in the time--like case. The $\bar r^{-1}$--part of the field is algebraically special (purely radiative) with degenerate principal null direction $\bar k^i$. It describes the electromagnetic radiation produced by the accelerated source just as in the time--like case. Unlike the time--like case the field here is singular at $\bar r=0$ (on the null world line of the source) \emph{and} at $\bar\xi ,\bar\eta \rightarrow\infty$, on account of (\ref{56}), which by (\ref{58}) corresponds to $\bar k^i$ pointing along the direction of the tangent $\bar v^i$ to the source world line.

Finally it is interesting to compare the model of an accelerated light--like charge described here with the model given by Synge~\cite{S1}. In our formalism Synge's potential 1--form is
\begin{align}
  \label{68}
  A=-e\,\bar h_0\,d\bar u\quad
  {\rm with}\quad
  \bar h_0=\bar a_i\,\bar k^i=\bar a^1(\bar u)\,\bar\xi+\bar a^2(\bar u)\,\bar\eta\ .
\end{align}
The corresponding Maxwell field and its dual are given by the 2--forms
\begin{align}
  \label{69}
  F=-\frac{e}{\bar r}\frac{\partial\bar h_0}{\partial\bar\xi}\,\bar\vartheta^1\wedge\bar\vartheta^4
  -\frac{e}{\bar r}\frac{\partial\bar h_0}{\partial\bar\eta}\,\bar\vartheta^2\wedge\bar\vartheta^4 =
  -\frac{e}{\bar r}\,\{\bar a^1\,\bar\vartheta^1\wedge\bar\vartheta^4
  +\bar a^2\,\bar\vartheta^2\wedge\bar\vartheta^4\}\ ,
\end{align}
and
\begin{align}\label{70}
{}^*F=\frac{e}{\bar r}\,\{-\bar a^1\,\bar\vartheta^2\wedge\bar\vartheta^4+\bar a^2\,\bar\vartheta^1\wedge\bar\vartheta^4\}=e\,\{-\bar a^1\,d\bar\eta\wedge d\bar u+\bar a^2\,d\bar\xi\wedge d\bar u\}\ .\end{align}
From the latter it is clear that Maxwell's vacuum field equations $d{}^*F=0$ are satisfied. If the world line $\bar r=0$ is a null geodesic then Synge's Maxwell field vanishes and there is no light--like analogue of the Coulomb field in his case. In terms of the coordinates $X^i$ the components $F_{ij}$ of the Maxwell field (\ref{69}) and the components 
${}^*F_{ij}$ of its dual (\ref{70}) read
\begin{align}\label{71}
F_{ij}=\frac{e}{\bar r}\left\{\left (\frac{\partial\bar h_0}{\partial\bar\xi}\frac{\partial\bar k_i}{\partial \bar\xi}+\frac{\partial\bar h_0}{\partial\bar\eta}\frac{\partial\bar k_i}{\partial \bar\eta}\right )\bar k_j-\left (\frac{\partial\bar h_0}{\partial\bar\xi}\frac{\partial\bar k_j}{\partial \bar\xi}+\frac{\partial\bar h_0}{\partial\bar\eta}\frac{\partial\bar k_j}{\partial \bar\eta}\right )\bar k_i\right\}\ ,\end{align}
and
\begin{align}\label{72}
{}^*F_{ij}=\frac{e}{\bar r}\left\{\left (\frac{\partial\bar h_0}{\partial\bar\xi}\frac{\partial\bar k_i}{\partial \bar\eta}-\frac{\partial\bar h_0}{\partial\bar\eta}\frac{\partial\bar k_i}{\partial \bar\xi}\right )\bar k_j-\left (\frac{\partial\bar h_0}{\partial\bar\xi}\frac{\partial\bar k_j}{\partial \bar\eta}-\frac{\partial\bar h_0}{\partial\bar\eta}\frac{\partial\bar k_j}{\partial \bar\xi}\right )\bar k_i\right\}\ ,\end{align}
respectively. From these we have
\begin{align}\label{73}
F_{ij}\,\bar k^j=0={}^*F_{ij}\,\bar k^j\ ,\end{align}
indicating that the Maxwell field is pure electromagnetic radiation with propagation direction $\bar k^i$ in Minkowskian space--time. When (\ref{63}) are substituted into (\ref{62}) 
using (\ref{61}) and raising the covariant index we have
\begin{align}\label{74}
\eta^{ij}=-\frac{\partial\bar k^i}{\partial\bar\xi}\frac{\partial\bar k^j}{\partial\bar\xi}-\frac{\partial\bar k^i}{\partial\bar\eta}\frac{\partial\bar k^j}{\partial\bar\eta}+\bar v^i\,\bar k^j+\bar v^j\,\bar k^i\ .\end{align}Hence with $\bar h_0$ given by (\ref{68}) we can write
\begin{align}\label{75}
\frac{\partial\bar h_0}{\partial\bar\xi}\frac{\partial\bar k_i}{\partial \bar\xi}+\frac{\partial\bar h_0}{\partial\bar\eta}\frac{\partial\bar k_i}{\partial \bar\eta}=\bar a^j\left (\frac{\partial\bar k_j}{\partial\bar\xi}\frac{\partial\bar k_i}{\partial\bar\xi}+\frac{\partial\bar k_j}{\partial\bar\eta}\frac{\partial\bar k_i}{\partial\bar\eta}\right )=\bar h_0\,\bar v_i-\bar a_i\ ,\end{align}
and so we can simplify (\ref{71}) to the form given originally by Synge:
\begin{align}\label{76}
F_{ij}=\frac{e}{\bar r}\left\{\bar k_i\,\bar a_j-\bar k_j\,\bar a_i+\bar h_0\,(\bar v_i\,\bar k_j-\bar v_j\,\bar k_i)\right\} 
\ .\end{align}From the first of (\ref{63}) Synge's potential 1--form in coordinates $X^i$ reads
\begin{align}\label{77}
A=-e\,\bar h_0\,\bar k_i\,dX^i=A_i\,dX^i\ .\end{align}Using the second of (\ref{63}) we can write the 4--potential here as
\begin{align}\label{78}
A_i=e\,\frac{\bar v_i}{\bar r}-e\,(\log\bar r)_{,i}\ ,\end{align}
demonstrating that, modulo a gauge term, the 4--potential has the same algebraic form as the Li\'enard--Wiechert 4--potential in the time--like case. In other words it is pointing in the direction of the tangent to the world line of the source.

Let us have a closer look at the electric and magnetic fields encoded in the Faraday tensor~(\ref{76}). To do so we introduce the notation $\vect{a} = a_{\alpha}$, $\alpha=1,2,3$, for the purely spatial part of the 4--vector $a_i$. It will prove useful to introduce the rescaled vector $\vect{K} = \bar{\vect{k}}/{\bar k_0}$. One can read off the electric field directly as $E_{\alpha} = F_{0\alpha}$ so that
\begin{align}
  \vect{E} &= \frac{e}{\bar r}
  \left\{
  \bar k_0(\bar{\vect{a}} - \bar h_0 \bar{\vect{v}}) -
  (\bar a_0 - \bar h_0 \bar v_0) \bar{\vect{k}} 
  \right\} =
  \frac{e}{\bar r}
  \left\{
  (\bar{\vect{a}} - \bar h_0 \bar{\vect{v}}) - (\bar a_0 - \bar h_0 \bar v_0) \vect{K}
  \right\} \bar k_0
  \ .
\end{align}
Likewise, we find for the magnetic field the expression
\begin{align}
  \vect{B} = \frac{e}{\bar r}
  \left\{
  (\bar{\vect{a}} - \bar h_0 \bar{\vect{v}}) \times \vect{k} 
  \right\} =
  \frac{e}{\bar r} \left\{
  (\bar{\vect{a}} - \bar h_0 \bar{\vect{v}}) \times \vect{K} 
  \right\} \bar k_0\ ,
\end{align}
which yields the somewhat expected relations
\begin{alignat}{2}
  \vect{E} \cdot \vect{B} &= 0\ , &\qquad
  |\vect{E}| = |\vect{B}| &= \frac{e}{\bar{r}} \sqrt{-\bar{a}_i\bar{a}^i}\, \bar{k}_0\ , \\
  \vect{E} \cdot \vect{K} &= 0\ , &\qquad \vect{E} \times \vect{K} &= \vect{B}\
  \qquad (\Rightarrow \vect{B} \cdot \vect{K} = 0).
\end{alignat}
This means we are dealing with an electromagnetic wave propagating along the direction defined by the spatial vector $\vect{K}$. Note that this vector is normalised to unity which follows from the fact that the original 4--vector $k_i$ was null. Hence these waves are travelling at the speed of light. As mentioned above, the electromagnetic field vanishes when the acceleration vanishes which corresponds to $\bar{r}=0$ being a geodesic world line. In~\cite{Kaz} a similar observation was made where $\bar{a}^i = d^2 w^i/du^2$ was interpreted as the curvature of the world line $w^i(u)$.

\section{Conclusions and discussions}

The model of a light--like charge with a non--geodesic world line given here places a great emphasis on geometry. The geometry involved utilizes a construction in the vicinity of an arbitrary light--like world line in Minkowskian space--time. This geometrical approach allows us to study charged particles moving with the speed of light directly. A key ingredient is the choice of parameter along such a world line since the natural parameter of arc length in the time-like case is not available in the light-like case. We have demonstrated in Section~\ref{sec:geo} the existence of a special parameter $\bar u$ along the light-like world line, unique up to a linear transformation, having the useful property in the current context that it becomes an affine parameter when the world line is a null geodesic. This appears to be a natural choice of parameter for the construction given in this paper which then allows us to construct the electric and magnetic fields of such a hypothetical particle. Our approach allows this construction without the need to introduce sources.

Interestingly Synge (private communication (1985)) has pointed out that `not quite obviously there exists on the curve a canonical parameter $u$ such that $a^ia_i=-1$ with $u$ defined to within an additive constant'. Synge's choice of parameter is very interesting and may be the natural choice for scenarios other than that considered here.

\end{document}